\documentclass[12pt]{article}
\usepackage{amstext}
\usepackage{graphicx}

\begin{document}

\date{}
\title{Magnetic influence on classical dispersion}
\author{T. N. C. Mendes$^{\star}$, Reinaldo de Melo e Souza$^{\dagger}$,
C. Farina$^*$\\
Instituto de F\'{\i}sica - UFRJ - CP 68528\\
Rio de Janeiro, RJ, Brasil - 21945-970.}
%
\maketitle

\begin{abstract}

We discuss the Lorentz model for dispersion and absorption of
radiation in dilute, linear and isotropic materials. Initially, with
the purpose of making the paper as self-contained as possible, we
reproduce the usual calculations concerning the interaction between
the charged material oscillators and the electric field of the
incident radiation, obtaining the main behavior of the reactive and
dissipative electromagnetic properties of the materials. Thereafter,
we also include the magnetic contribution of the Lorentz force to
the equation of motion of the oscillators up to first order in
$v/c$, which leads to some interesting results, like the
approximately linear dependence of the refraction index with the
radiation intensity and the appearance of a second region of
anomalous dispersion around  half the natural frequencies of the
material.
\end{abstract}

\bigskip
\vfill \noindent $^{\star }$ {e-mail: tarciro@if.ufrj.br}

\noindent $^\dagger$ {e-mail: reinaldo@if.ufrj.br}

\noindent $^*$ {e-mail: farina@if.ufrj.br}

\pagebreak

\section{Introduction}

As described by A. Pais, \cite{Pais}, Kirchoff's 1859 pioneering
work about the search of the spectral function of thermal radiation
raised the discussion of the interaction between matter and
radiation, an issue that, since then, has shown to be of great
importance in many branches of physics, like quantum field theory,
condensed matter physics and optical physics, to mention just a few.
In this work, we discuss the Lorentz model, one of its most
prodigious child. Born in the last years of the 19-th century, this
model treats matter as posessing stable electronic charge
configurations, held together by harmonic potentials, and that
interact with electromagnetic fields of a given radiation. This
model achieves amazing results describing optical and dissipative
properties of material media (for instance, their dielectric
constants, magnetic permeabilities and absorption/dispersion
factors), specially those with low refractive index. Usually, when
discussing classical dispersion theory, only the electric
contribution is considered since this contribution is $c/v$ times
larger than the magnetic one. These issues are often covered by
undergraduation textbooks in classical electromagnetism
\cite{Griffiths,Marion}, but are discussed in more advanced
textbooks as well \cite{Jackson2nd}.

Here, we will show that the inclusion of the usually ignored
magnetic term leads to quite interesting results as, for instance,
the appearence of a second region of anomalous dispersion. We also
present some numerical analysis in order to see if experimental
verifications of the new effects discussed in this paper are
plausible, particulary for electromagnetic radiation of high enough
intensity.

We finish this brief introduction with a comment to motivate the
search for other possibilities of ocurrence of anomalous dispersion.
Besides their intrinsic importance, it has been known for a long
time that group velocity in regions of anomalous dispersion close to
an absorption line may exceed the speed of light in vacuum and,
under some circunstances, may become infinite or even negative
\cite{BrillouinBook1960,ChiaoPA1993}. For a recent and pedagogical
discussion on negative group velocity see \cite{McDonaldAJP2001} and
references therein.

\section{Lorentz's model}       

For future convenience, and to make this article as self-contained
as possible, we start with a brief review of the Lorentz model for
the electric permittivity $\epsilon (\omega)$ of linear and
isotropic media. The interaction between charges and fields in this
model is given by the Lorentz force. For the $j$-th  charged
particle with negative electric charge $-e_j$ whose position at
instant $t$ is denoted by $\mathbf{r_{j}}$,  this force is given, in
gaussian units, by
\begin{equation}
\mathbf{F}_j(\mathbf{r_{j}},t)=-e_j\left[\mathbf{E}(\mathbf{r_{j}},t)
 + \frac{\mathbf{v}_j}{c}\times \mathbf{B}(\mathbf{r_{j}},t)\right]
\end{equation}
where $\mathbf{E}$ and $\mathbf{B}$ are, respectively, the electric
and magnetic fields of the electromagnetic wave, $c$ is the velocity
of light  in vacuum and $\mathbf{v}_j = {\dot{\bf r}}_j$ is the
velocity of the $j$-th particle at instant $t$. Denoting by
$\mathbf{x}_j$ the difference between $\mathbf{r}_j$ and the
equilibrium position of the particle, denoted by
$\mathbf{r_{j\,0}}$, we are lead to the following equation of motion
for such a particle
\begin{equation}
\label{EqMov}
\mathbf{\ddot x}_j+2\gamma_j\mathbf{\dot x}_j+\omega_j^2\mathbf{x}_j
 = -{e_j\over m_j}\left[\mathbf{E}_j(t) +
 \frac{\mathbf{\dot x}_j}{c}\times \mathbf{B}_j(t)\right]
\end{equation}
where, in order to simplify the notation, $\mathbf{E}_j(t)$ and
$\mathbf{B}_j(t)$ stand for the electric and magnetic fields at
instant $t$ and position $\mathbf{r_{j0}}$, that is,
$\mathbf{E}_j(t)=\mathbf{E}(\mathbf{r_{j\, 0}},t)$ and
$\mathbf{B}_j(t)=\mathbf{B}(\mathbf{r_{j\, 0}},t)$, and the
quantities $\gamma_j$
 and $\omega_j$ are the damping constant and the natural oscillation frequency
 associated to the $j$-th  charged particle. In
(\ref{EqMov}), we have assumed that $\vert\mathbf{x}_j\vert$ is
small enough so that the spatial variations of the electromagnetic
fields over a distance of the order of the atom diameter can be
neglected ($\vert\mathbf{x}_j\vert\ll  \lambda$, with $\lambda$
being the wavelength of the incident electromagnetic
 wave). In other words, the electromagnetic
 fields can be considered as if they were uniform in space for each
 charged partice (this is the so called dipole approximation). In
 fact, since we will need to talk about the polarization ${\bf P}$
 and magnetization ${\bf M}$ of the material, which are macroscopic
 quantities defined as volumetric densities computed over a volume
 $\delta V$ which is very small macroscopically but big enough to
 contain many thousands of atoms, we shall assume that
 the condition $\vert\mathbf{x}_j\vert^3\ll \delta V/\ll
 \lambda^3$ is valid.

For simplicity, we consider a linearly polarized and monochromatic
plane wave of  angular frequency $\omega$ and choose the cartesian
axis in such a way that the electromagnetic fields are written as
\begin{equation}
 \mathbf{E}_j(t) =  E_{\omega}\cos(\omega t - \alpha_j)
  \,\mathbf{\hat y}\;;\;\;\;\;\;
\mathbf{B}_j(t) = E_{\omega}
 \cos(\omega t - \alpha_j)
 \,\mathbf{\hat z}
\;;\;\;\;\;\;\mathbf{k}=k\,\mathbf{\hat x}\, ,
 \end{equation}
 where $\alpha_j$ is a constant, $k=\omega/c$, $\mathbf{\hat x}$, $\mathbf{\hat y}$ and $\mathbf{\hat
z}$ constitute a right-handed orthonormal set and $E_{\omega}$ is
the amplitude of the electromagnetic wave at position
 $\mathbf{r}_{j0}$. It depends on  $\mathbf{r}_{j0}$, but this fact was not
 indicated explicitly to mantain the notation as simple as possible.
 Since the time averages to be taken do not depend on $\alpha_j$, we
 shall take $\alpha_j=0$ without loss of generality (alternatively, we can get rid off
 $\alpha_j$ with a simple time translation $t\;\rightarrow\; t +
 \alpha_j/\omega$).  With this in mind, equation (\ref{EqMov}) takes the form (in components)
\begin{eqnarray}
\label{eqx}
\ddot x_j+2\gamma_j\dot x_j+\omega_j^2 x_j &=&
 -{\dot y_j\over c}{e_j E_{\omega}\over m_j}\cos\omega t
\\
%
\label{eqy}
\ddot y_j+2\gamma_j\dot y_j+\omega_j^2 y_j &=&
 -\left(1-{\dot x_j\over c}\right){e_jE_{\omega}\over m_j}\cos\omega t
\\
%
\label{eqz}
\ddot z_j+2\gamma_j\dot z_j+\omega_j^2 z_j&=&0
\end{eqnarray}
where $\mathbf{x}_j = x_j\mathbf{\hat x}+y_j\mathbf{\hat y} +
z_j\mathbf{\hat z}$. Looking for stationary solutions only, we see
that the motion of the particle is restricted to the ${\cal OXY}$
plane, perpendicular to the magnetic field.
 Restricting to non-relativistic cases, that is,
   assuming $\vert\dot x_j\vert\ll c$ and $\vert \dot y_j\vert\ll c$,
   equations (\ref{eqx}) and (\ref{eqy}) can be approximated, respectively, by
\begin{equation}
\ddot x_j+2\gamma_j\dot x_j+\omega_j^2 x_j = 0\;
\;\;\;\;\;\mbox{and}\;\;\;\;\;
 \ddot y_j+2\gamma_j\dot y_j+\omega_j^2 y_j=-{e_j E_{\omega}\over m_j}\cos\omega t\,.
 \end{equation}
 Hence, at this order, the stationary solutions of equations
  (\ref{eqx}), (\ref{eqy}) and (\ref{eqz}), are
\begin{equation}
\label{Soly}
x_j(t) = 0\; ;\;\;\;\;\;
 y_j(t)=A_{\omega}\cos\left(\omega t+\phi_{\omega}\right)\; ;
 \;\;\;\;\;\;
 z_j(t)=0\, ,
\end{equation}
in which  $A_{\omega}={{\mathcal E}_j/Z_{\omega}}$ and
\begin{equation}
\label{Zfi}
{\mathcal E}_j=-{e_j E_{\omega}\over m_j}\;,\;\;Z_{\omega}
 = \left[\left(\omega_j^2-\omega^2\right)^2+4\gamma_j^2\omega^2\right]^{1/2}
 \!\!,\;\;\cos\phi_{\omega}={\omega_j^2-\omega^2\over
Z_{\omega}}\;,\;\;\;\sin\phi_{\omega}=-{2\gamma_j\omega\over
Z_{\omega}}\, .
\end{equation}
Observe that the  particle oscillates in the same direction of the
electric field with the same angular frequency $\omega$, but shifted
by a phase $\phi_{\omega}$.

\begin{figure}[!h]
\begin{center}
\includegraphics[width=4.0in]{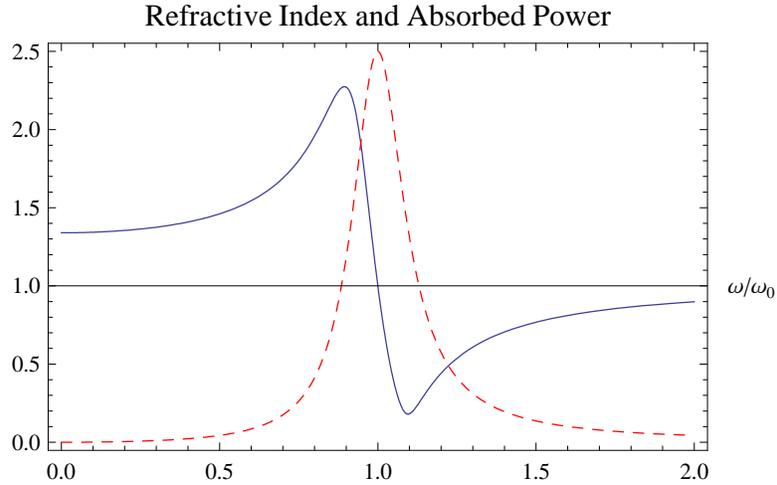}
\caption{Refractive index $\, n(\omega)=\sqrt{\epsilon(\omega)}\,$
(solid line) and
 absorbed power $\,{\mathcal P}(\omega)\,$ (dashed line), for fixed
intensity $I_{\omega}$ as functions of the field frequency, for a
medium with only one natural frequency $\omega_0$ (for convenience,
we used the ratio
 $\omega/\omega_0$). In both graphs we used
a  natural width $\gamma_0=0,\!1\omega_0$.
The values on the vertical axis are written in arbitrary units.}
\label{NP}
\end{center}
\end{figure}

Suppose that in a region with a volume $\delta V$ (where the field
variation is negligible by assumption) there are $n_j=N_j\delta V$
oscillators with charge $-e_{j}$, mass $m_{j}$ and angular frequency
$\omega_j$. Therefore, the total number of oscillators in $\delta V$
is then $N_{\delta V}=\sum_jn_j=\sum_jN_j\delta V$ and the
polarization due to the present field is
\begin{equation}
\mathbf{P} = -\sum_j N_je_j\mathbf{x}_j\, .
 \end{equation}
 The mean energy per volume in a period, {\it reversibly} stored in the medium, is
\begin{equation}
\label{Up1}
U_P=-\langle\mathbf{P}\cdot\mathbf{E}\rangle =
 E_{\omega}\sum_j {N_je_j{{\mathcal E}_j}\over Z_{\omega}}\langle \cos\left(\omega t
 + \phi_{\omega}\right)\cos\omega t\rangle =
 -{1\over 2}E_{\omega}^2\sum_j{N_je_j^2\over m_j}
 \left({\omega_j^2-\omega^2\over Z_{\omega}^2}\right)\, .
\end{equation}
On the other hand, this same energy may be written as
\begin{equation}\label{Up2}
U_P = -{1\over 2}{\bar P}_{\omega}E_{\omega} =
 -{1\over 2}\chi^{\,\prime}({\omega})E^2_{\omega}\, ,
 \end{equation}
 where we used that ${\bar P}_{\omega} =
 \chi^{\,\prime}\left(\omega\right)E_{\omega}$ for the Fourier
 transform of the polarization, with $\chi^{\,\prime}(\omega)$ being the
 electric susceptibility of the medium. Comparing equations (\ref{Up1}) and (\ref{Up2}), we
 identify
\begin{equation}
  \chi^{\,\prime}\left(\omega\right)
 = \sum_j{N_j\alpha_j\omega_j^2
 \left(\omega_j^2-\omega^2\right)\over \left(\omega_j^2-\omega^2\right)^2
 + 4\gamma_j^2\omega^2}\, ,\label{ChiEletrico}
\end{equation}
where $\alpha_j=e_j^2/m_j\omega_j^2$ is the static polarizability of
the $j$-th oscillator. Comparing last equation to the following
relation between polariazation and electric field in a linear and
isotropic material (see, for instance, Chapter 4 of Ref.
\cite{Jackson2nd})
 \begin{equation}
\mathbf{P}(\omega)={3\over 4\pi}\left(\epsilon(\omega)-1\over
\epsilon(\omega)+2\right)\mathbf{E}(\omega)\, ,
 \end{equation}
 we obtain
\begin{equation}
\label{Eps}
\epsilon(\omega)={3+8\pi\chi^{\,\prime}\left(\omega\right)\over
3-4\pi\chi^{\,\prime} \left(\omega\right)}
 ¨\;\;\;\Longrightarrow\;\;\;
 \epsilon(\omega)\simeq 1 + 4\pi\chi^{\,\prime}\left(\omega\right)\,,
\end{equation}
where $\epsilon(\omega)$ is the dielectric \lq\lq constant" of the
material (it is not a constant since it is a function of $\omega$).
Last approximation, $\chi^{\,\prime}\left(\omega\right)\ll 1$, is
valid specially for low density materials, as gases, for instance.
Considering non-magnetic materials and neglecting the magnetic
contribution, we have $\mu(\omega)=1$, which makes the refractive
index $n(\omega)$ approximately equal to
\begin{equation}
\label{nEps}
n(\omega)=\sqrt{\epsilon(\omega)}\simeq 1+2\pi\chi^{\,\prime}\left(\omega\right)\,.
\end{equation}
The behavior of $\; n(\omega)=\sqrt{\epsilon(\omega)}\;$ is shown in
Figure \ref{NP} for a single natural frequency $\omega_0$ dominant
in the medium. Between $\omega\simeq\omega_0-\gamma_0$ and
$\omega\simeq\omega_0+\gamma_0$ we see a region of anomalous
dispersion ($dn/d\omega<0$).

The mean power per volume absorbed by the medium is given by
\begin{equation}
\label{PotM}
{\mathcal P}(\omega)=-\sum_j N_je_j\langle\mathbf{\dot x}_j\cdot\mathbf{E}\rangle = -\sum_j {N_je_j^2\over m_j Z_{\omega}}\,\omega E_{\omega}^2\langle\sin\left(\omega t+\phi_{\omega}\right)\cos\omega t\rangle = \sum_j {N_je_j^2\over m_j}\,{\gamma_j\omega^2\over Z_{\omega}^2}\,E_{\omega}^2\,.
\end{equation}
 In terms of the mean value of the radiation intensity $I_{\omega}$,
 the above equation takes the form
\begin{equation}
\label{PotI}
{\mathcal P}(\omega)={8\pi\over c}\,\omega\chi^{\,\prime\prime}(\omega)I_{\omega}
 \;\;,\;\;\;\chi^{\,\prime\prime}(\omega) =
 \sum_j{N_j\alpha_j\omega_j^2\gamma_j\omega\over\left(\omega_j^2-\omega^2\right)^2
 + 4\gamma_j^2\omega^2}\; ,
 \;\;\;\;\; \alpha_j = \frac{e^2_j}{m_j \omega_j^2}\,,
\end{equation}
reflecting the intuitive fact that the power is proportional to the
intensity.

In Figure \ref{NP} we show the behavior of (\ref{PotI}) as a
function of the angular frequency with a fixed intensity. We
considered also that the medium has only one dominating natural
frequency $\omega_0$. We see that ${\mathcal P}(\omega)$ is very
small at all frequencies except those in the anomalous-dispersion
region. Specially, for $\omega=\omega_0$ the electric field is in
resonance with natural oscillators of the medium and the power
absorbed is maximum.

%
%

\section{Magnetic contribution to dispersion}

In this section, we shall  repeat the previous calculations but
without neglecting the influence of the magnetic field. However, we
shall compute only the first order correction (in order
$\vert{\dot{\bf x}}_j\vert/c$) to the stationary solutions written
in (\ref{Soly}). As it will become evident in a moment, to obtain
the new stationary corrections to these equations we need to obtain
only the new stationary solution for the component $x_j(t)$, namely,
to consider the previously neglected $\dot y_j/c$ term on the rhs of
equation (\ref{eqx}). Hence, it suffices to substitute
 $y_j(t) = A_\omega\,\cos(\omega t +\phi_\omega)$ into equation (\ref{eqx}),
which leads to
\begin{eqnarray}
\ddot x_j+2\gamma_j\dot x_j+\omega_j^2 x_j
 &=&
 -\Bigl[ -\omega A_\omega\,\sin(\omega t + \phi_\omega)\Bigr]
 \frac{e_jE_\omega}{m_j\, c}\cos(\omega t)\cr\cr
 &=&
 -\frac{{\cal E}^2_j\omega}{cZ_\omega}
 \,\sin(\omega t+\phi_\omega) \cos(\omega t)
  \cr\cr
 &=&
 -\frac{{\cal E}^2_j\omega}{2cZ_\omega}
\Bigl[\,\sin(2\omega t + \phi_\omega) +
 \sin\phi_\omega\Bigr]\cr\cr
 &=&
 {\gamma_j{\mathcal E}_j^2\over c}{\omega^2\over
Z_{\omega}^2}\;-\;{{\mathcal E}_j^2\over 2c}{\omega\over
Z_{\omega}}\sin\left(2\omega t+\phi_{\omega}\right)\, ,
\end{eqnarray}
where we used the definitions $A_\omega = {\cal E}_j/Z_\omega =
-e_jE_\omega/(m_j Z_\omega)$, the trigonometric identity $\sin a\cos
b = (1/2)[\sin(a+b) + \sin(a-b)]$ and that
$\sin\phi_\omega=-2\gamma_j\omega/Z_\omega$.

A simple inspection on the rhs of the previous differential equation
allows us to state that its stationary solution will oscillate with
angular frequency $2\omega$, rather than $\omega$. Following the
usual procedure of obtaining stationary solutions of damped harmonic
oscillators under an external force given by a constant term plus a
harmonic one, we can obtain in a straightforward way the expression
for the stationary solution for $x_j(t)$. For convenience, we write
below also the stationary solution for $y_j(t)$ up to first order in
$\vert{\dot{\bf x}}_j\vert/c$:
\begin{eqnarray}
\label{xj}
x_j(t) &=& {\gamma_j\over \omega_j^2}\,\omega kA_{\omega}^2 -
 {1\over 2}kA_{\omega}A_{2\omega}\sin\left(2\omega
t+\phi_{\omega}+\phi_{2\omega}\right)\\
 y_j(t) &=& A_\omega \cos(\omega t + \phi_\omega)\, ,\label{yjB}
\end{eqnarray}
where $k=\omega/c$ and the quantities $A_{2\omega}$, $Z_{2\omega}$
and $\phi_{2\omega}$ can be obtained from equation (\ref{Zfi})
 by just substituting $\omega$ by $2\omega$. As we shall see,
 equation (\ref{yjB}) is, indeed, the correct expression up to this
 order. But before justifying this statement, a few comments are in
 order:

 \noindent
 {\it (i)} the stationary solution for $z_j(t)$ remains
 $z_j(t) = 0$, so that, as mentioned before, the stationary motion occurs
 on the ${\cal OXY}$ plane, which is perpendicular to the magnetic
 field;

\noindent
 {\it (ii)}  while the motion of the charged particle along the
 ${\cal OY}$ direction is an oscillation around an equilibrium position
 at $y_j=0$, its motion along the ${\cal OX}$ direction is an
 oscillation around the equilibrium position
 $x_j = \mbox{\large$\frac{\gamma_j}{\omega_j^2}$}\,\omega kA_{\omega}^2$,
which depends on the frequency of the electromagnetic fields. This
dependence is similar  as that of the absorbed power by the medium,
namely: it is zero for $\omega=0$ or $\omega \rightarrow\infty$ and
has its maximum at $\omega = \omega_j$;

\noindent
 {\it (iii)} comparing the amplitudes of oscillation of equations
 (\ref{xj}) and (\ref{yjB}) we see that
the former is $kA_{2\omega}$ times the latter. Hence, the
oscillations in the direction of the electric field (${\cal OY}$
direction) are greater than those along the  direction of the wave
vector $\mathbf{k}$ (${\cal OX}$ direction).

Let us now estimate the error committed by neglecting the term $\dot
x_j/c$ in (\ref{eqy}). Substituting (\ref{xj}) into (\ref{eqy}), and
writing the solution of the resulting differential equation as
$y_j^1(t) = y_j^{0}(t) + \Delta y_j(t)$, where $y_j^{0}(t)$ is given
by expression (\ref{Soly}), it can be shown that $\Delta y_j(t)$ is
given by
\begin{equation}
\Delta y_j(t)=-{1\over 4}k^2A_{\omega}^2A_{2\omega}\cos\left(\omega
t+2\phi_{\omega}+\phi_{2\omega}\right)\;-\;{1\over
4}k^2A_{\omega}A_{2\omega}A_{3\omega}\cos\left(2\omega
t+\phi_{\omega}+\phi_{2\omega}+\phi_{3\omega}\right)\, ,
 \end{equation}
 where the quantities $A_{3\omega}$ and $\phi_{3\omega}$ can be
 determined in the same way as $A_{2\omega}$ and $\phi_{2\omega}$,
 described before. Looking at the last equation, and remembering
 that $k^2=\omega^2/c^2$, it is not difficult to conclude that
 $\Delta y_j(t)$ is of order $\vert{\dot{\bf x}}_j\vert^2/c^2$ and
 hence it can be neglected if we maintain terms only ut to first
 order in $\vert{\dot{\bf x}}_j\vert/c$. That is why $\Delta y_j(t)$
 is not present in equation (\ref{yjB}).

\begin{figure}[!h]
\begin{center}
\includegraphics[width=4.2in]{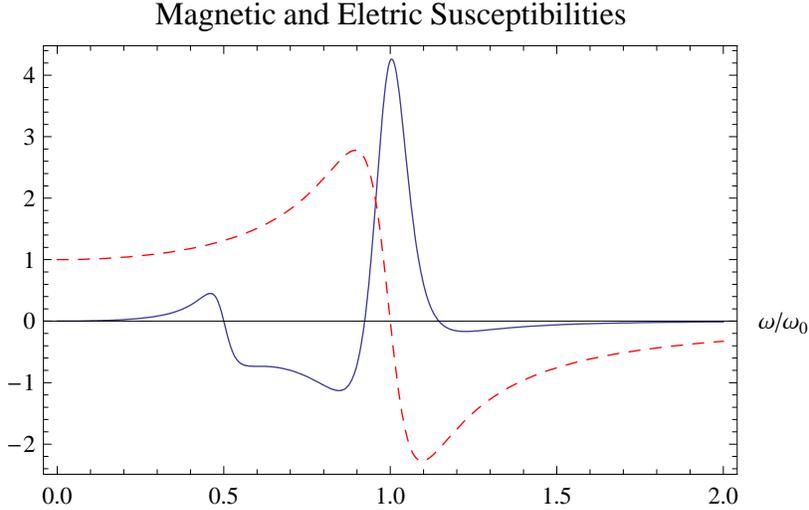}
\caption{Eelectric and magnetic susceptibilities, given by equations
(\ref{Up2}) and (\ref{Xm}) as functions of the ratio
$\omega/\omega_0$. The dashed line stands for
$\chi^{\,\prime}(\omega)$ while the solid line corresponds to
$\chi_m(\omega)$. For simplicity, we considered only one natural
frequency, $\omega_0$, (and the corresponding natural line width
$\gamma_0$). We have made $\gamma_0/\omega_0 = 0.1$ in both graphs.
The values on the vertical axis are written in arbitrary units.}
\label{Chi}
\end{center}
\end{figure}

The magnetic dipole momentum of the $j$-th oscillator is given by
\begin{eqnarray}
\label{mm}
\mathbf{m}_j
 &=&
 -{e_j\over 2c}\left(\mathbf{x}_j\times \mathbf{\dot x}_j\right)\cr\cr
 &=&
 -{e_j\over 2c}\left(x_j{\dot y}_j-y_j{\dot x}_j\right)\mathbf{\hat z}
\cr
&=& -\;{1\over 4}\, e_j\mathcal E_jk^2A_{\omega}A_{2\omega}\,
 \mathbf{\hat z}\bigg[\cos(\omega t+\phi_{\omega})\cos(2\omega
t+\phi_{\omega}+\phi_{2\omega})\;+
\cr
&{\;}& \; +\;\;\cos(\omega
t+\phi_{2\omega})-{2\gamma_j\omega\over\omega_j^2}
 {Z_{2\omega}\over Z_{\omega}}\sin(\omega t+\phi_{\omega})\bigg]\,,
\end{eqnarray}
so that the corresponding medium magnetization is
\begin{equation}
\mathbf{M} = \sum_j N_j\mathbf{m}_j
 \end{equation}
 and the medium mean energy per volume stored in the medium reads
\begin{eqnarray}
\label{Mj}
U_m &=&-\langle\mathbf{M}\cdot\mathbf{B}\rangle\cr\cr
 &=&
 -\sum_j {N_je_j^4 E_{\omega}^4\over 4m_j^3 c^2}
 {\omega^2\over Z_{\omega}^2Z_{2\omega}}
 \left[{3\over 4}\cos\phi_{2\omega} -
 {\gamma_j\omega\over \omega_j^2}{Z_{2\omega}\over Z_{\omega}}\sin\phi_{\omega}\right]
\cr\cr
&=&-\sum_j {N_je_j^4 E_{\omega}^4\over 4m_j^3 c^2} {\omega^2\over
Z_{\omega}^2}\left[{3\over 4} \left({\omega_j^2-4\omega^2\over
Z_{2\omega}^2}\right) + {1\over 2}\left({2\gamma_j\omega\over
\omega_j Z_{\omega}}\right)^2\right]\, .
\end{eqnarray}
Following a procedure totally analogous to that employed in Section
{\bf 2} for the electric polarization, it is possible to write the
magnetic energy as
\begin{equation}
\label{Mmedio}
U_m=-{1\over 2}\bar
M_{\omega}E_{\omega}\;\;\;\Longrightarrow\;\;\;\bar M_{\omega}
 =\sum_j {3 N_je_j^4 E_{\omega}^3\over 8m_j^3 c^2}
 {\omega^2\over Z_{\omega}^2}\left[{\omega_j^2-4\omega^2\over Z_{2\omega}^2} +
 {2\over 3}\left({2\gamma_j\omega\over \omega_j Z_{\omega}}\right)^2\right]\, ,
\end{equation}
where $\mathbf{M}=\bar M_{\omega}\,\mathbf{\hat z}$ is
 the mean induced magnetization parallel to the magnetic field.
Comparing last equation to the relation between magnetization and
the magnetic induction ${\bf B}$ in a linear medium, namely, (see,
for instance, Chapter 5 of Ref. \cite{Jackson2nd})
\begin{equation}
{\bf M}(\omega) = \frac{3}{4\pi} \left(\frac{\mu(\omega) -1}
 {\mu(\omega) + 2}\right) {\bf B}\, ,
\end{equation}
we obtain, after some algebraic rearrangements,
\begin{eqnarray}
\label{mu}
\mu(\omega) &=& {3+8\pi\chi_m\left(\omega\right)\over
3-4\pi\chi_m\left(\omega\right)}
 \;\;\;\Longrightarrow\;\;\;
\mu(\omega)\simeq 1+4\pi\chi_m\left(\omega\right)\,,
\\
%
\label{Xm}
\chi_m(\omega)  \!\!\!\!&=& \!\!\!\!\! \sum_j \!\!{3\pi
N_j\beta_j\omega_j^4\omega^2I_{\omega} \over
\left(\omega_j^2-\omega^2\right)^2+4\gamma_j^2\omega^2}
\!\left[\!\!{\omega_j^2-4\omega^2\over
\left(\omega_j^2-4\omega^2\right)^2+16\gamma_j^2\omega^2} \!+\!
{8\gamma_j^2\omega^2/3\omega_j^2\over
\left(\omega_j^2-\omega^2\right)^2+4\gamma_j^2\omega^2}\!\right]
\end{eqnarray}
In the last equation, we defined
$\beta_j=e_j^4/m_j^3c^3\omega_j^4=\alpha_j^2/m_j c^3$ and one could,
in principle, think of  $\chi_m(\omega)$ as the magnetic
susceptibility of the medium to the magnetic field, something
analogous to the electric susceptibility $\chi^{\,\prime}(\omega)$
defined by equation (\ref{Up2}). Note that $\chi_m(\omega)$ has
 a dependence on the frequency of the electromagnetic wave which is
 much more complicated than that exhibited by the electric
 susceptibility $\chi^{\,\prime}(\omega)$, as it is
 illustrated in Figure \ref{Chi}.

 Differently from $\chi^{\,\prime}(\omega)$, which varies smoothly with the
 frequency  except in the range of anomalous dispersion (with a
width $2\gamma_0$ centered at $\omega_0$), $\chi_m(\omega)$ shows an
anomalous behavior around $\omega=\omega_0/2$. There is also a
pronounced peak around $\omega=\omega_0$, similar to that appearing
in the absorbed power by the medium, as shown in Figure \ref{NP},
but whose intensity varies with $1/\gamma_0^2$ instead of
$1/\gamma_0$.

\begin{figure}[!h]
\begin{center}
\includegraphics[width=4.2in]{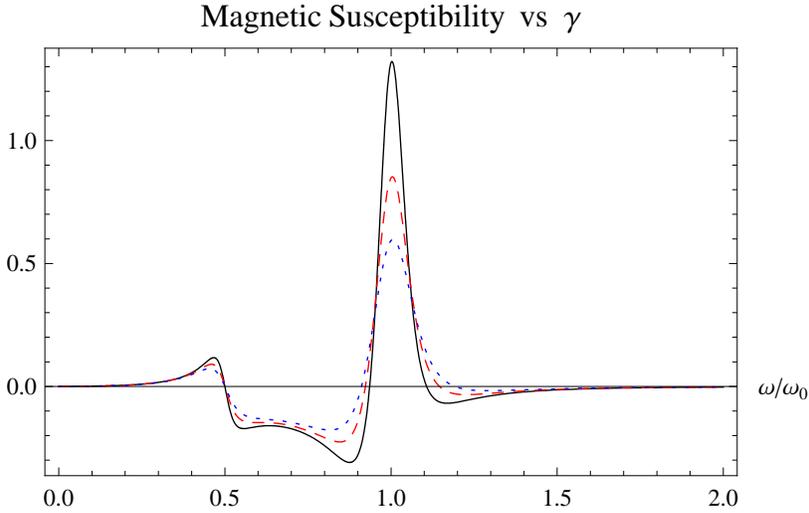}
\caption{Magnetic susceptibility $\chi_m(\omega)$ as a function of
the ratio $\omega/\omega_0$  for different values of $\gamma_0$. The
solid line corresponds to $\gamma_0/\omega_0=0.08$, while the dashed
line, to $\gamma_0/\omega_0=0.1$, and the dotted line, to
$\gamma_0/\omega_0=0.12$. The values on the vertical axis are in
arbitrary units.}
\label{MS}
\end{center}
\end{figure}

In Figure \ref{MS} we plot some graphs for $\chi_m(\omega)$ {\it
versus} $\omega/\omega_0$ for different values of $\gamma_0$. In the
range ${1\over 2}\omega_0<\omega<\omega_0$ we observe an uncommon
behavior with the frequency, while for high frequencies,
$\omega\gg\omega_0$, $\chi_m(\omega)$ goes to zero faster than
$\chi^{\,\prime}(\omega)$. It is worth mentioning that, for low
frequencies, $\omega\ll\omega_0$, only $\chi_m(\omega)$ goes to
zero.

In contrast to the electric permittivity $\epsilon(\omega)$, given
by equation (\ref{Eps}), the magnetic permeability $\mu(\omega)$
depends on the intensity $I_\omega$ of the electromagnetic wave, as
can be seen from equations (\ref{mu}) and (\ref{Xm}). For
$\chi_m(\omega)\ll 1$, which is generally the case, $\mu(\omega) -
1$ is proportional to $I_\omega$. Including the magnetic
contribution, the refractive index takes the form
\begin{equation}
\label{nI}
n(\omega) = \sqrt{\epsilon(\omega)\mu(\omega)}
 \;\;\;\;\Longrightarrow\;\;\;\;
n(\omega)\simeq
1+2\pi\Big[\chi^{\,\prime}(\omega)+\chi_m(\omega)\Big]\, .
\end{equation}
\begin{figure}[!h]
\begin{center}
\includegraphics[width=4.0in]{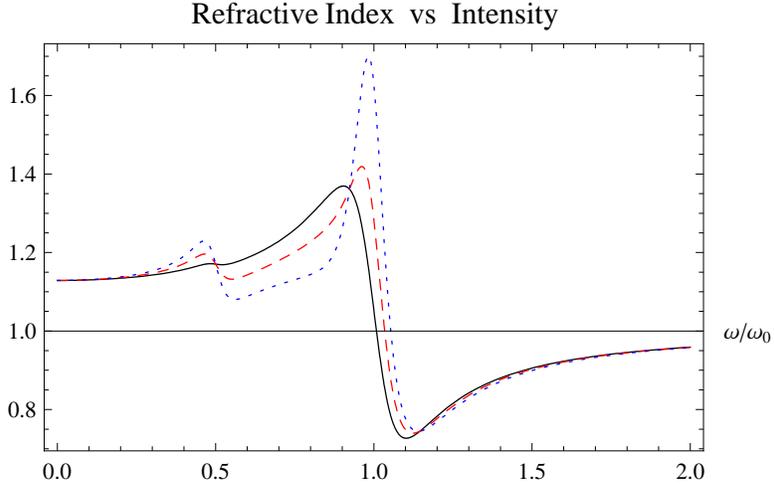}
\caption{Refractive index, given by equation (\ref{nI}), for
different intensities of the incident electromagnetic wave and
$\gamma_0/\omega_0=0.1\omega_0$. The curve with solid line
corresponds to an intensity  $5$ times smaller than the intensity
used to draw the curve with dashed line and $10$ times smaller than
that used in the curve with dotted line. The values on the vertical
axis are written in arbitrary units.}
\label{NW}
\end{center}
\end{figure}
\begin{figure}[!h]
\begin{center}
\includegraphics[width=4.4in]{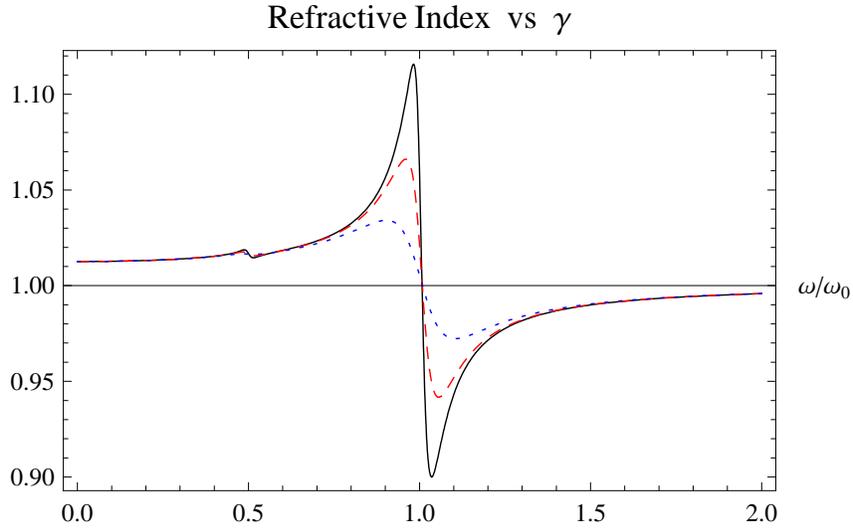}
\caption{Refractive index, given by (\ref{nI}), as a function of
$\omega/\omega_0$ for different values of $\gamma_0$ and fixed
intensity. The solid line stands for $\gamma_0/\omega_0=0.03$, the
 dashed line, to $\gamma_0/\omega_0=0.05$ and the dotted line, to
$\gamma_0/\omega_0=0.1$. The values on the vertical axis are in
arbitrary units.}
\label{NWG}
\end{center}
\end{figure}

We see that the role of the magnetic field is to sum
$2\pi\chi_m(\omega)$ to the rhs of equation (\ref{nEps}), which
introduces a linear dependence with the intensity of the
electromagnetic wave for fixed frequencies. This term, however,
 is usually much smaller than the electric contribution, but may become
relevant for values of intensity that make
$\chi^{\,\prime}(\omega)/\chi_m(\omega)\sim 1$. In the next section
we discuss if there is a possibility of checking experimentally this
 magnetic contribution. Figure \ref {NW} shows the effect of this
new magnetic term on the expression for the refractive index.

For low values of the intensity $I_\omega$, but still high enough to
make the contribution due to  $\chi_m(\omega)$ a relevant one, the
frequency dependence of the refractive index is very similar to that
shown in Figure \ref{NP}, except for the appearence of a range of a
secondary anomalous dispersion (SAD), centered at $\omega={1\over
2}\omega_0$ (besides the usual anomalous dispersion around
$\omega=\omega_0$). This fact may be viewed as a first signature of
the influence of the intensity on the refractive index.

As the intensity increases, the behavior of $n(\omega)$, in the
interval ${1\over 2}\omega_0<\omega<\omega_0$, becomes more and more
similar to that of $\chi_m(\omega)$. We also notice that the
intersection between the curve for $n(\omega)$ and the line
$n(\omega)=1$, which occurs at $\omega=\omega_0$ for vanishing
intensity, is shifted as the intensity is increased. Finally, for
$\omega\ll\omega_0$ and $\omega>\omega_0$, $n(\omega)$  does not
change appreciably when one varies the intensity.

 In Figure \ref{NWG} we  see the relation between  $n(\omega)$
  and frequency for different values of $\gamma_0$, keeping the
  incident intensity fixed. As expected, as $\gamma_0$ becomes smaller and smaller,
  the secondary anomalous dispersion becomes more and  more evident in
  the interval $\omega={1\over 2}\omega_0$.

\section{Numerical estimatives}

 So far we have discussed the problem of the interaction between
  microscopic structures and electromagnetic radiation until
  first order in $v/c$, without worrying about numerical values
  that could sustain the possibility of experimental observation
  of those predictions  (magnetic corrections to the
  reactive properties of substances). In this section, we want to analyze
   the limits for the parameters associated to the radiation (frequency
    and intensity) and to the system (natural frequencies) that make
    such predictions observable within the validity domain of this model.

Henceforth, we consider our material oscillators as electrons with
electric charge $e=-4,\!803\cdot
10^{-10}\left(\mathrm{erg}\cdot\mathrm{cm}\right)^{1/2}$ and mass
$m=9,\!109\cdot 10^{-28}\mathrm{g}$. We shall call them actives
since they are the only charges that can interact with the incident
radiation. For the sake of simplicity, we shall assume that there is
 only one natural frequency present, denoted by $\omega_0$, so every
electron oscillates with the same frequency. Therefore, the sum in
(\ref{Up2}), (\ref{PotI}) and (\ref{Xm}) keeps only one term. We
also admit that these oscillators are perfect harmonic oscillators,
an assumption that leads to
\begin{equation}
\gamma_0\equiv{e^2\omega_0^2\over 3mc^3}\, .
 \end{equation}
Strictly speaking, this model is applicable only to substances
formed by active electrons that behave as harmonic oscillators.
However, it can be successfully extended to those that actually are
not formed by oscillators, but in these cases the parameters
$\alpha_j$, $\beta_j$, $\omega_j$ and $\gamma_j$ have to be
considered independent from each other, being adjusted through
experimental data.

Let us, then, analyse the secondary anomalous dispersion. In the
range where it occurs, we can approximate the magnetic
susceptibility $\chi_m(\omega)$ by
\begin{equation}
\label{XDAS}
\chi_m(\omega)\simeq {2\pi\over 3}N_0\beta_0\omega_0I_{w}
{\omega_0-2\omega\over\left(\omega_0-2\omega\right)^2+\gamma_0^2}\,
,
\end{equation}
while, in this range, the electric susceptibility is almost
constant, and can be taken as being
\begin{equation}
\chi^{\,\prime}(\omega)\simeq {4\over 3}N_0\alpha_0\, .
 \end{equation}
 We have seen that magnetic effects will become relevant for
$\vert\chi^{\,\prime}(\omega)/\chi_m(\omega)\vert\sim 1$. The
maximum value of $\vert\chi_m(\omega)\vert$ occurs around ${1\over
2}\left(\omega_0\pm\gamma_0\right)$ and is given by
\begin{equation}
\vert\chi_m\vert\simeq{\pi\over
3}N_0\beta_0I_{\omega}{\omega_0\over\gamma_0}={\pi N_0e^2\over
m^2\omega_0^5}\, ,
 \end{equation}
  so that the minimum intensity needed is
\begin{equation}
\label{Imin}
I_{\mathrm{min}}\sim{{4\over 3}N_0\alpha_0\over{\pi\over
3}N_0\beta_0\omega_0/\gamma_0}={32\pi^2\over
3}{mc^3\over\lambda_0^3}\simeq
260\left({\mu\mathrm{m}\over\lambda_0}\right)^3\mathrm{GW}/\mathrm{cm}^2\,.
\end{equation}
These intensities are relatively small in the microwave region while
they are very high in the optical region: for
$\lambda_0=0,\!5\mu\mathrm{m}$ the intensity is something like
$2,\!0\mathrm{TW/cm^2}$, making it hard to observe the secondary
anomalous dispersion in that range of frequencies. Nevertheless, in
the range from infrared to microwaves, an experimental observation
of these magnetic effects seems feasible: for
$\lambda_0=100\mu\mathrm{m}$, for instance, $I_{\mathrm{min}}\sim
260\,\mathrm{kW/cm^2}$, a value that can be achieved  nowadays by
lasers.

%
\section{Conclusions and final remarks}

In this work we have discussed the Lorentz model for dispersion and
absorption of electromagnetic radiation in  diluted, linear and
isotropic material media. We started reviewing the usual computation
of the electric permittivity of the material that describes the
response of the material oscillators only to the electric field of a
linearly polarized and monocromatic electromagnetic wave. In this
simplified model,  the magnetic field is not taken into account,
since its contribution is already of order $v/c$. From this model,
one obtains the main results about the electromagnetic reactive and
dissipative properties of matter, namely: the dependence of the
refractive index with the radiation frequency (dispersion); the
existence of a region of strong anomalous dispersion for the
frequency range $\vert\omega_j-\omega\vert\sim\gamma_j$ (which we
call, for convenience, a primary anomalous dispersion), where
$dn/d\omega<0$; the absorbed/dissipated power is proportional to the
intensity of radiation and varies quickly in the range
$\vert\omega_j-\omega\vert\sim\gamma_j$, achieving its maximum value
at resonance.

Then we discussed what are the first corrections to the previous
model when we include the magnetic term of the Lorentz force into
the equations of motion of the material oscillators.
 Despite this term does not change substantially
  the dissipative properties of the medium, it leads, when
solved up to first order approximation in $v/c$, to quite
interesting (and presumably observable) effects concerning the
reactive properties of matter, namely: {\it (i)} the refractive
index acquires a dependence on the intensity of the electromagnetic
wave (besides its dependence on the wave frequency); {\it (ii)} the
appearance of a secondary zone of anomalous dispersion in the region
$\mbox{$\vert{1\over 2}\omega_j-\omega\vert\sim\gamma_j$}$ and of an
intensification of the primary dispersion  peak. This result breaks
the monotonic increasing behavior of the refractive index for
sufficiently high intensities, in the range ${1\over
2}\omega_j<\omega<\omega_j$;

The observation of the new magnetic effects just described in the
resonance region ($\omega\sim\omega_j$) is frustrated due to the
strong energy absorption. However, the situation may not be the same
in the secondary anomalous dispersion zone, where absorption is very
small, so that the required high intensities of the incident
radiation may be achieved. In order to investigate the plausibility
of making real experiments with the present technology with the
purpose of observing the secondary anomalous dispersion, we have
presented in the previous section some numerical estimatives.
 The observation of this secondary anomalous dispersion
would be a signature of the dependence of the refractive index on
the intensity of the electromagnetic wave.

As a final comment, we would like to say that, so far, we have
considered only a linearly polarized incident radiation. However,
one can use circularly polarized radiation as well. In this case,
though the magnetization acquires a non-vanishing component along
the direction of propagation of the electromagnetic wave, the final
results remain the same, since the direction of propagation is
perpendicular to the magnetic field, so that the contribution of
this extra term vanish in the expression of
 $\langle{\bf M}\cdot{\bf B}\rangle$.

{

\end{document}